\documentclass[a4paper, 12pt]{article}

\author{Clovis Jacinto de Matos\footnote{ESA-HQ, European Space Agency, 8-10 rue Mario Nikis, 75015 Paris, France, e-mail:
Clovis.de.Matos@esa.int}}

\title{Gravitoelectromagnetism and Dark Energy in Superconductors}

\begin{document}

\maketitle

\begin{abstract}
A gravitomagnetic analogue of the London moment in superconductors
can explain the anomalous Cooper pair mass excess reported by
Janet Tate. Ultimately the gravitomagnetic London moment is
attributed to the breaking of the principle of general covariance
in superconductors. This naturally implies non-conservation of
classical energy-momentum. Possible relation with the
manifestation of dark energy in superconductors is questioned.
\end{abstract}


\section{Introduction}

In 1989 Tate et \emph{al.}\cite{Tate89} \cite{Tate90} reported a
Cooper pair mass excess in Niobium of 84 parts per million greater
than twice the free electron mass ($m_e$), whereas a theoretical
calculation based on the theory of relativity predicts a value 8
parts per million less than $2m_e$. This disagreement between
theory and experiment has not been resolved so far
\cite{Liu}$^-$\cite{Capelle}. Conjecturing an additional
gravitomagnetic term in the Cooper pairs canonical momentum
accounts for Tate's observations. This naturally leads to the
conjecture that Tate's excess of mass is not real but instead a
rotating superconductor would simply exhibit a gravitomagnetic
analogue to the well known magnetic London moment. However the
magnitude of this conjectured gravitomagnetic field would be 10
orders of magnitude higher than Earth natural gravitomagnetic
field (about $10^{-14} Rad/s$).

The electromagnetic properties of superconductors can be explained
through the breaking of gauge symmetry and consequently through a
massive photon in the superconductive material. A similar
mechanism  for gravitation involving the breaking of the Principle
of General Covariance (PGC) in superconductors would lead to a set
of Proca type equations for gravitoelectromagnetism, with an
associated massive spin one boson to convey the
gravitoelectromagnetic interaction. Requiring that the PGC is
recovered from this set of equations in the case of normal matter;
we find back for the case of physical systems made simultaneously
of coherent and normal matter, like superconductors, the
anomalously high gravitomagnetic London moment conjectured from
Tate's experiment. Ultimately It appears that Tate's measurements
can be expressed in terms of the ratio between the Copper pair
mass density, $\rho^*_m$, and the superconductor's bulk density,
$\rho_m$.
\begin{equation}
\frac{m^*-m}{m}\simeq\frac{\rho^*_m}{\rho_m}\label{equ0}
\end{equation}
Where $m^*$ and $m$ are respectively the experimentally measured
and the theoretically predicted Cooper pair mass.

The breaking of the PGC, implies that energy-momentum would not be
conserved in superconductors. Could this be related with some
still unknown properties of dark energy? The investigation of the
physical nature of dark energy in quantum materials is a
fascinating possibility, which Beck and Mackey\cite{Beck} are
already exploring for the case of Josephson junctions.

\section{Gravitomagnetic London Moment}

Tate et \emph{al.} used a SQUID to measure the magnetic field
generated by the rotation ($\omega=2\pi\nu [Rad/s]$) of a thin (on
the order of the London penetration depth) Niobium superconductive
ring, also called the London moment. Following Ginzburg-Landau
theory of superconductivity the total magnetic flux (including the
Cooper pairs current density) cancels at regular frequency
intervals $\Delta \nu [s^{-1}]$.
\begin{equation}
\frac{\hbar}{m^*}=2S\Delta\nu\label{equ1}
\end{equation}
Where $S$ is the area bounded by the Niobium ring. Based on Eq.
(\ref{equ1}) Tate estimated the Cooper pair's mass,
$m^*=1.82203\times10^{-30} [Kg]$, which is higher than the
theoretically expected $m=1.82186\times10^{-30} [Kg]$, including
relativistic corrections\cite{Cabrera}.

In order to assess Tate's experiment, following DeWitt and Ross, a
gravitomagnetic term is added to the Copper pair's canonical
momentum\cite{DeWitt} \cite{Ross}, $\vec{p_s}$.
\begin{equation}
\vec{p_s}=m\vec{v_s}+e\vec{A}+m\vec{A_g}\label{equ2}
\end{equation}
where $e$ is the Cooper pair's electric charge, $\vec{v_s}$ is the
Cooper pair's velocity, $\vec{A}$ and $\vec{A_g}$ are respectively
the magnetic and gravitomagnetic vector potentials.

Using again Ginzburg-Landau theory we find that the zero-flux
condition, Eq. (\ref{equ1}), depends now also on a gravitomagnetic
component\cite{Kiefer} \cite{Bertolami}.
\begin{equation}
\frac{\hbar}{m}=2S\Big(\Delta \nu + \frac{\Delta
B_g}{4\pi}\Big)\label{equ3}
\end{equation}
Doing Eq.(\ref{equ1}) into Eq. (\ref{equ3}) we find the
gravitomagnetic variation needed to account for Tate's anomalous
excess of mass\cite{Tajmar03} is: $\Delta B_g=1.65\times10^{-5}
[Rad/s]$.

Subtracting the classical Cooper pair's canonical momentum:
$\vec{p^*_s}=m^*\vec{v_s}+e\vec{A}$, from the generalised one, Eq.
(\ref{equ2}), and taking the rotational, we obtain the
gravitomagnetic analogue of the London moment in superconductors,
which we call the "Gravitomagnetic London Moment"\cite{Tajmar05}.
\begin{equation}
\vec{B_g}=\frac{m^*-m}{m}2\vec{\omega}=\frac{\Delta m}{m}2
\vec{\omega} \label{equ4}
\end{equation}
Therefore we conclude that the Cooper pair's mass does not
increase but instead when a superconductor is set rotating it
generates simultaneously a (homogeneous) magnetic field (London
Moment) together with an (homogeneous) gravitomagnetic field
(Gravitomagnetic London Moment). If the latest phenomenon is
neglected it is naturally interpreted, as Tate did, as being an
anomalous excess of mass of the Cooper pairs.

The gravitomagnetic field generated by the rotating Niobium ring
in Tate's experiment would then be:
\begin{equation}
\vec{B_g}=1.84\times10^{-4}\vec{\omega},\label{equ5}
\end{equation}
which is very large even for small angular velocities compared to
classical astronomical sources, but cannot be ruled out based on
the experimental results achieved so far\cite{Tajmar06}. However
how can we explain this conjecture? We will see that the answer
passes through the investigation of the validity of the Principle
of General Covariance for superconductors.

\section{Spontaneous Breaking of Gauge Invariance in Superconductors}

Superconductor's properties (zero resistivity, Meissner effect,
London moment, flux quantization, Josephson effect etc...)can be
understood likewise spontaneous breaking of electromagnetic gauge
invariance when the material is in the superconductive
phase\cite{Ryder} \cite{Weinberg1}. In field theory, this symmetry
breaking leads to massive photons via the Higgs mechanism. In this
case the Maxwell equations transform to the so-called
Maxwell-Proca equations, which are given by:
\begin{eqnarray}
\nabla \vec{E} =\frac{\rho^*}{\epsilon_0}-\frac{1}{\lambda_\gamma
^2}\phi \label{equ6}\,, \\
\nabla \vec{B}=0 \label{equ7}\,,\\
\nabla\times \vec{E}=-\dot{\vec{B}}\label{equ8}\,,\\
\nabla\times\vec{B}=\mu_0 \rho^* \vec{v_s}+\frac{1}{c^2}\dot{\vec
E}-\frac{1}{\lambda_\gamma^2}\vec{A}, \label{equ9}
\end{eqnarray}
Where $\vec{E}$ is the electric field, $\vec{B}$ is the magnetic
field, $\epsilon_{0}$ is the vacuum electric permittivity,
$\mu_{0}=1 /\epsilon_0  c^2$ is vacuum magnetic permeability,
$\phi$ is the scalar electric potential, $\vec{A}$ is the magnetic
vector potential, $\rho^*$ is the Cooper pairs fluid electric
density, $\vec{v_S}$ is the cooper pairs velocity, and
$\lambda_\gamma=\hbar/m_\gamma c$ is the photon's Compton
wavelength, which is equal to the London penetration depth
$\lambda_L=\sqrt{\frac{m}{\mu_0\rho^* e}}$.

Taking the rotational of Eq. (\ref{equ9}) and neglecting the term
coming from the displacement current, we get the following
equation for the magnetic field:
\begin{equation}
\nabla^2\vec{B}+\frac{1}{\lambda_\gamma^2}\vec{B}=\frac{1}{\lambda_L^2}\frac{m}{e}2\vec{\omega}.\label{equ10}
\end{equation}
Solving Eq. (\ref{equ10}) for a one dimensional case, we obtain
the Meissner effect and the London moment.
\begin{equation}
B=B_0 e^{-x/{\lambda_\gamma}}+2\omega
\frac{m}{e}\Big(\frac{\lambda_\gamma}{\lambda_L}\Big)^2\label{equ11}
\end{equation}
Following the argument from Becker et \emph{al.}\cite{Becker} and
London \cite{London}, the London moment is developed by a net
current that is lagging behind the positive lattice matrix, The
Cooper pair current density direction sign has to show in opposite
direction than the angular velocity of the superconducting bulk.
This is important as the London moment in all measurements, due to
the negative charge of the Cooper pair, shows in the same
direction as the angular velocity. Having
$\lambda_\gamma=\lambda_L$ we finally get:
\begin{equation}
B=B_0 e^{-x/{\lambda_L}}-2\omega \frac{m}{e}\label{equ11}
\end{equation}

\section{Spontaneous Breaking of the Principle of General Covariance in
Superconductors}

General Relativity is founded on the \emph{principle of
equivalence}, which rests on the equality between the inertial and
the gravitational mass of any physical system, and formulates that
\emph{at every space-time point in an arbitrary gravitational
field it is possible to choose a "locally inertial coordinate
system" such that, within a sufficiently small region of the point
in question, the laws of nature take the same form as in
unaccelerated Cartesian coordinate systems in the absence of
gravity}. In other words, The inertial frames, that is, the
"freely falling coordinate systems", are indeed determined by the
local gravitational field, which arises from all the matter in the
universe, far and near. However, once in an inertial frame, the
laws of motion are completely unaffected by the presence of nearby
masses, either gravitationally or in any other way.

Following Steven Weinberg, the \emph{Principle of General
Covariance} (PGC) is an alternative version of the principle of
equivalence\cite{Weinberg}, which is very appropriate to
investigate the field equations for electromagnetism and
gravitation. It states that \emph{a physical equation holds in a
general gravitational field, if two conditions are met}:
\begin{enumerate}
\item The equation holds in the absence of gravitation; that is, it
agrees with the laws of special relativity when the metric tensor
$g_{\alpha\beta}$ equals the Minkowsky tensor $\eta_{\alpha\beta}$
and when the affine connection $\Gamma_{\beta\gamma}^{\alpha}$
vanishes.
\item The equation is generally covariant; that is, it preserves
its form under a general coordinate transformation $x \rightarrow
x'$.
\end{enumerate}

It should be stressed that general covariance by itself is empty
of physical content. The significance of the principle of general
covariance lies in its statement about the effects of gravitation,
that a physical equation by virtue of its general covariance will
be true in a gravitational field if it is true in the absence of
gravitation. The PGC is not an invariance principle, like the
principle of Galilean or special relativity, but is instead a
statement about the effects of gravitation, and about nothing
else. In particular general covariance does not imply Lorentz
invariance. Any physical principle such as the PGC, which takes
the form of an invariance principle but whose content is actually
limited to a restriction on the interaction of one particular
field, is called a dynamic symmetry. As discussed above local
gauge invariance, which governs the electromagnetic interaction is
another important dynamical symmetry. We can actually say that the
Principle of General Covariance in general relativity is the
analogous of the Principle of Gauge Invariance in electrodynamics.

Maxwell-Proca equations for electromagnetism
Eqs.\ref{equ6}-\ref{equ9}, which apply in a superconductor, are
not gauge invariant just as they are not generally covariant. If
we assume that the PGC is spontaneously broken in a
superconductor, like gauge invariance is, the weak field
approximation of Einstein field equations would lead to the
following set of Proca equations for gravitoelectromagnetism,
which contains a spin 1 massive boson, called graviphoton, to
convey the gravitoelectromagnetic interaction \cite{Argyris}
\cite{de Matos}.
\begin{eqnarray}
\nabla \vec{g} =-\frac{\rho^*_m}{\epsilon_{0g}}-\frac{1}{\lambda_g
^2}\phi_g \label{equ12}\,, \\
\nabla \vec{B_g}=0 \label{equ13}\,,\\
\nabla\times \vec{g}=-\dot{\vec{B_g}}\label{equ14}\,,\\
\nabla\times\vec{B_g}=-\mu_{0g} \rho^*_m
\vec{v_s}+\frac{1}{c^2}\dot{\vec
g}-\frac{1}{\lambda_g^2}\vec{A_g}, \label{equ15}
\end{eqnarray}
Where $\vec{g}$ is the gravitational field, $\vec{B_g}$ is the
gravitomagnetic field, $\epsilon_{0g}=1/4 \pi G$ is the vacuum
gravitational permittivity, $\mu_{0g}=4\pi G / c^2$ is vacuum
gravitomagnetic permeability, $\phi_g$ is the scalar gravitational
potential, $\vec{A_g}$ is the gravitomagnetic vector potential,
$\rho^*_m$ is the Cooper pairs mass density, $\vec{v_S}$ is the
cooper pairs velocity, and $\lambda_g=\hbar/m_g c$ is the Compton
wavelength of the graviphoton.

Taking the gradient of Eq. (\ref{equ12}), and the rotational of
Eq. (\ref{equ15}), and solving the resulting differential
equations for the one dimensional case we find respectively the
form of the principle of equivalence and of the gravitomagnetic
Larmor Theorem\cite{Mashhoon} in superconductive cavities.
\begin{eqnarray}
\vec g=-\vec a \mu_{0g} \rho^*_m \lambda^2_g\label{equ16}\,,\\
\vec B_g= 2\vec\omega\mu_{0g} \rho^*_m \lambda^2_g \label{equ17}
\end{eqnarray}
where for Eq. (\ref{equ17}) we had to introduce Becker's argument
that the Cooper pairs are lagging behind the lattice so that the
current is flowing in the opposite direction of $\omega$.

In order to find a phenomenological law for the graviphoton
wavelength\cite{Tajmar06_2} we request that from Eq. (\ref{equ16})
and (\ref{equ17}) the PGC is restored in the case of normal
matter. In that case the mass density reduces to the materials
bulk density, $\rho_m$, no condensate phase is present within the
material.
\begin{eqnarray}
\vec g=-\vec a \mu_{0g} \rho_m \lambda^2_g\label{equ18}\,,\\
\vec B_g= 2\omega\mu_{0g} \rho_m \lambda^2_g \label{equ19}
\end{eqnarray}
Since normal matter complies with the PGC, Eqs. (\ref{equ18}) and
(\ref{equ19}) must reduce to:
\begin{eqnarray}
\vec g=-\vec a\label{equ20}\,,\\
\vec B_g= 2\vec\omega\label{equ21}
\end{eqnarray}
Comparing Eq. (\ref{equ20}) and (\ref{equ21}) with Eq.
(\ref{equ18}) and (\ref{equ19}) we find that the gravitphoton
Compton wavelength must be inversely proportional to the local
density of the bulk material mass:
\begin{equation}
\frac{1}{\lambda^2_g}=\mu_{0g}\rho_m\label{equ22}
\end{equation}
Doing Eq. (\ref{equ22}) into Eq. (\ref{equ16}) and (\ref{equ17})
we get:
\begin{eqnarray}
\vec g=-\vec a \frac {\rho^*_m}{\rho_m}\label{equ23}\,,\\
\vec B_g= 2\vec\omega \frac  {\rho^*_m}{\rho_m} \label{equ24}
\end{eqnarray}
which clearly indicate a breaking of general covariance. Notice
that in the case of Bose Einstein Condensates (BEC) we have only
one single condensate phase in our material ($\rho^*_m=\rho_m$)
implying that the PGC is no violated in BECs.

Comparing Eq. (\ref{equ24}) with Eq. (\ref{equ4}) and (\ref{equ5})
we can explain the additional gravitomagnetic term in the Cooper
pairs canonical momentum, in function of the mass density ratio of
the coherent and normal phase.
\begin{equation}
9.2\times10^{-5}=\frac{m^*-m}{m}\simeq\frac{\rho^*_m}{\rho_m}=3.9\times10^{-6}\label{equ25}
\end{equation}
The numerical values in Eq. (\ref{equ25}) correspond to the case
on Niobium in Tate's experimental conditions.

\section{Dark Energy in Superconductors?}

It is well known that breaking of general covariance leads to
non-conservation of energy-momentum (in the covariant
sense)\cite{Weinberg}. Would that mean that in a superconductor we
could observe some manifestation of dark energy?

Presently the physical nature of dark energy is unknown. What is
clear is that various astronomical observations (supernovae, CMB
fluctuations, large-scale structure) provide rather convincing
evidence that around 73 percent of the energy contents of the
universe is a rather homogeneous form of energy, so-called "dark
energy". A large number of theoretical models exist for dark
energy, but an entirely convincing theoretical breakthrough has
not yet been achieved. Popular models are based on quintessence
fields, phantom fields, quintom fields, Born-Infeld quantum
condensates, the Chaplygin gas, fields with non-standard kinetic
terms, possible links between the cosmological constant and the
graviton mass \cite{Tajmarcosmo} to name just a few (see for
example Refs. \cite{Peebles} and \cite{Padmanabhan} for reviews).
All of these approaches contain "new physics" in one way or
another, though at different levels. However, it is clear that the
number of possible dark energy models that are based on new
physics is infinite. Only experiments will ultimately be able to
confirm or refute the various theoretical constructs\cite{Beck2}.

Beck is currently exploring the possibility that vacuum
fluctuations, allowed by the uncertainty relation, in Josephson
junction create dark energy. This is a priori the simplest
explanation for dark energy. Assuming that the total vacuum energy
density associated with zero-point fluctuations cannot exceed the
presently measured dark energy density of the universe, Beck
predicts an upper cutoff frequency of $\nu_c=(1.69\pm 0.05)\times
10^{12} Hz$ for the measured frequency spectrum of the zero point
fluctuations in the Josephson junction. The largest frequencies
that have been reached in the experiments are of the same order of
magnitude as $\nu_c$ and provide a lower bound on the dark energy
density of the universe. If this is confirmed by future
experiments how would that be related with the photon and
graviphoton mass in superconductors? Where does the mass of these
particles come from in the superconductive material? Finally Can
we break general covariance in a superconductor without violating
energy momentum conservation, if a superconductor contains a new
form of energy?

\section{Conclusion}

The close analogy between the Principle of General Covariance and
Gauge invariance, allows us to investigate the
gravitoelectromagnetic properties of quantum materials in the
framework of massive gravitoelectromagnetic Proca equations. We
find that the breaking of the PGC in superconductors leads to a
gravitomagnetic London moment and an associated additional
gravitomagnetic term in the Cooper pairs canonical momentum, which
can explain the anomalous excess of mass of Cooper pairs reported
by Tate.

The breaking of the PGC in superconductors implies the
non-equivalence between a rigid reference frame made with
superconductive walls (superconductive cavity), being uniformly
accelerated in a gravitational field free region,
\begin{eqnarray}
\vec g=-\vec a \Big(1+ \frac {\rho^*_m}{\rho_m}\Big)\label{equ26}\,,\\
\vec B_g= 2\vec\omega \Big( 1+ \frac {\rho^*_m}{\rho_m}\Big)
\label{equ27}
\end{eqnarray}
and a classical rigid reference frame (made with normal matter) in
a similar situation.
\begin{equation}
\vec g=-\vec a, \hspace{1 cm} \vec B_g= 2\vec\omega \label{equ28}
\end{equation}
However breaking the principle of general covariance leads to a
violation of the law of conservation of energy-momentum. It is not
clear yet if this would be a sign for some manifestation of dark
energy in superconductive materials. However it worths further
investigation.

\section{Acknowledgement}

I am grateful to Prof. Orfeu Bertolami, Prof. John Moffat, Prof.
Francis Everitt, Prof. Alan Kostelecky, for fruitful discussions
and pertinent comments. I would like also to thank Prof. Raymond
Chiao for encouragements and for being a rich source of
inspiration to the present work. My profound gratitude also goes
to Dr. Slava Turyshev and to all the organizers of the "From
Quantum to Cosmos" conference for the fantastic forum of debates,
they succeeded to establish during the conference.


\end{document}